\documentstyle[12pt,epsfig]{article}
\def \ins#1#2#3#4#5#6#7 {
  \begin{figure}[#6]
    \hskip #5
    \special{em:graph {./pcx/#2} }
    \vskip #4
    \caption{#7}
    \label{#1}
  \end{figure}}

\begin{document}

\begin{center}
{\Large \bf
Manifestation of 12-Quark Bag State of ${}^4He$ Nucleus in Elastic
$d~{}^4He$ Scattering }
\end{center}

\begin{center}
A.M. Mosallem \footnote{Math. \& Theor. Phys. Dept., NRC,
AEA, Cairo, Egypt.}, V.V.Uzhinskii
\end{center}

\begin{center}
Laboratory of Information Technologies\\
Joint Institute for Nuclear Research, \\
Dubna, Russia
\end{center}
\vspace{1cm}
%------------------------- ABSTRACT ----------------------------------
\noindent
The ${}^4He~d$ elastic scattering at the momentum of 19.8 GeV/c is
analyzed in the framework of the Glauber theory. The scattering
amplitude was evaluated using different sets of values of the
nucleon-nucleon amplitude parameters and the ${}^4He$ density function
as a superposition of the Gaussian functions. It is shown that
it is impossible to describe simultaneously the $p~{}^4He$ and
$d~{}^4He$ elastic scattering cross sections using the same set of the
$NN$-amplitude parameters. Inclusion of the twelve quark bag
admixture to the ground state of the ${}^4He$ nucleus in the
calculations allows one to reproduce the experimental data quite well.
It is shown that the admixture manifests itself in the $d~{}^4He$
elastic scattering in all region of the momentum transfer. At small $t$
the effect can be at the level of $\sim$ 10\%. At large $t$ it can be
$\sim$ 30\%.

\newpage
\setcounter{page}{1}

%*********************************************************************
\section*{Introduction}
%*********************************************************************
Understanding how the nuclei are react with other nuclei is one of the
fundamental goals of nuclear physics. Elastic and inelastic
nucleus-nucleus scattering has been studied for years to extract
information on the nuclear structure and different aspects of the
nuclear reactions. The other interesting topic is a manifestation of
the quarks in the reactions. In high energy interactions quarks and
gluons received high transverse momentum are observed as jets of
hadrons. At lower energies the jet production cross section becomes
extraordinary small, and the jets have not been registered at
experimental studies until now. At the same time it is known that
reactions with participation of the ${}^4He$ nucleus can not be
described quite well within the framework of standard nuclear physics.
In paper by L.G.Dakno and N.N.Nikolaev \cite{Nik-Dakhno} it was assumed
and shown that $12\%$ admixture of twelve quark bag configuration in
the ground state wave function of the ${}^4He$ nucleus allows one to
understand the irregularities of proton -${}^4He$ elastic scattering at
high energies.  We returned to the hypothesis in our previous paper
\cite{Ahm-Uzh}, and have shown that it really gives an opportunity to
describe $p~{}^4He$-scattering in a wide energy range. In present paper
we continue our study, and consider $d~{}^4He$ elastic scattering.

Experimental data on the $d~{}^4He$ elastic scattering at the
laboratory momentum of 19.8 GeV/c were presented in Ref. \cite{Avde}.
They were analyzed in the framework of the Glauber theory
\cite{Glauber, Inoz}.  The key quantities of the theory are
characteristics of $NN$ elastic scattering amplitude and
parametrization of the ground state wave function of the ${}^4He$
nucleus. In Ref. \cite{Avde} it was chosen the simplest gaussian
parametrization of the wave function. The $NN$ characteristics were
considered as the fitting parameters. As a result, all of these hid a
big discrepancy between the experimental data and calculations with real
parameters. We show it in the next Sec. where the main technical
details of our calculations are given. Inclusion of the twelve quark
bag state of the ${}^4He$ nucleus in the calculation scheme and
manifestation of the state in the elastic scattering is considered in
Sec. 2. In the last Sec. we summarize our results.

%*********************************************************************
\section{Calculation of $d~{}^4He$ elastic cross section}
%*********************************************************************

The Glauber amplitude of nucleus-nucleus
scattering has a form \cite{Franco, Czyz-Max, Hansen}:
\begin{equation}                                 %**************eq1
F_{AB}(\vec q)=\frac{ip}{2\pi}\int d^2b~ e^{i\vec q \cdot \vec b}
\Gamma\left(\vec b \right),
\label{eq1}
\end{equation}                                   %**************eq1
\begin{equation}                                 %**************eq2
\Gamma\left(\vec b \right)= \langle\psi_A^f\psi_B^f |1-\prod_{j=1}^A
\prod_{k=1}^B \left(1- \gamma(\vec b - \vec s_j+\vec
\tau_k)\right)|\psi_A^i\psi_B^i\rangle,
\label{eq2}
\end{equation}                                   %**************eq2
where $\vec b$ is the impact parameter, $p$ is the momentum of
the projectile nucleus, $\psi_A^i,\psi_B^i$ and $\psi_A^f,\psi_B^f$
are the initial and final states wave functions of the projectile and
the target nucleus, respectively,  $\gamma$ is the $NN$ elastic
scattering amplitude in the impact parameter representation. In high
energy physics it is often parametrized as:
\begin{equation}                                  %**************eq3
\gamma(\vec b)=\beta~e^{-\vec b^2/2B_{NN}},
\label{eq3}
\end{equation}                                    %**************eq3
\noindent where $\beta= \sigma_{NN}^{tot}\left(1-i\alpha_{NN}\right)
/\left(4\pi B_{NN}\right)$, $\sigma_{NN}^{tot}$ is the $NN$ total cross
section, $B_{NN}$ is the slope parameter of the $NN$ differential
elastic cross section at zero momentum transfer, $\alpha_{NN}$ -- the
ratio of the real to imaginary parts of the $NN$ elastic scattering
amplitude at zero momentum transfer.

The elastic nucleus-nucleus differential cross section is
determined  as
\begin{equation}                                 %**************eq4
\frac{d\sigma}{d\Omega} = \left|F_{AB}\right|^2.
\label{eq4}
\end{equation}                                   %**************eq4

To take into account all terms of the expansion of the product in Eq.
(\ref{eq2}), one can represent each term like that shown in the Fig. 1,
where the circles correspond to the interacting nuclei, the black
and white points -- to the nucleons, the solid lines -- to the
interactions between nucleons.
%-----------------------------------------------------------------------
%-----------------------------Fig1--------------------------------------
%-----------------------------------------------------------------------
\begin{figure}[cbth]
\begin{center}
\psfig{file=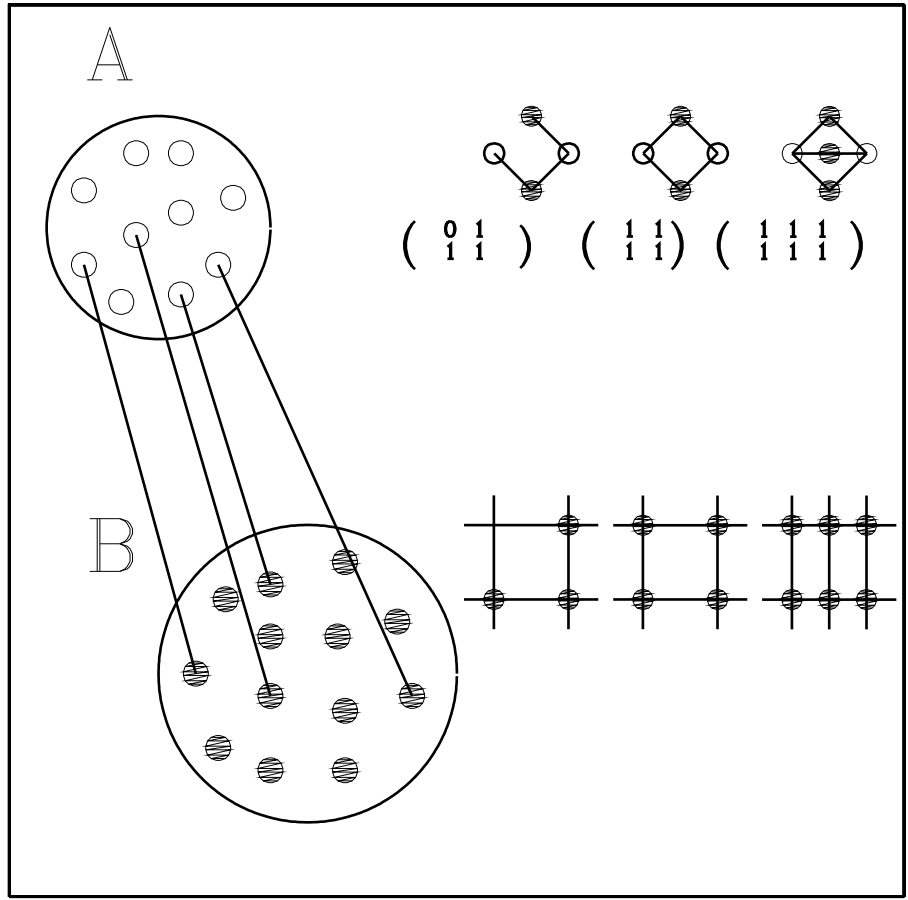,width=120mm,height=120mm,angle=0}
\caption{Graphical representation of the multiple scattering terms}
\end{center}
\end{figure}
%-----------------------------------------------------------------------
%-----------------------------------------------------------------------
%-----------------------------------------------------------------------

Using the diagrams, one can calculate how many the diagrams of whatever
type are. It is pure combinatorical problem which can be solved with a
help of the graph theory.  In the graph theory the diagrams of the Fig.
1 are called bi-colored labelled graphs. This graphs can be represented
with the help of an adjacency matrix. The adjacency matrix
$\underline{\underline{D}}=\left[d_{ij}\right]$ of a labelled graph $G$
is a matrix of order $A\times B$ in which $d_{ij}=1$ if points $i$ and $j$
are adjacent (are connected with a line) and $d_{ij}=0$ in other case.
By the other way, the graph can be represented by the set of crossing
points of A horizontal and B vertical lines with dark circles in the
places corresponding to the elements $d_{ij}=1$. This representation is
called net graph representation. We will refer to the net graphs as the
scattering diagrams.

Each term in the expansion of Eq. (\ref{eq2}) has a form

\begin{equation}                                 %**************eq5
-\langle\psi_A^f\psi_B^f |\prod_{(j,k)}\left(-
\gamma(\vec b - \vec s_j+\vec \tau_k)\right)|\psi_A^i\psi_B^i\rangle,
\label{eq5}
\end{equation}                                   %**************eq5
$$(j,k)\in M \subset \{I_A\}\otimes\{I_B\}$$ $$
\{I_A\}=(1,2,3,...,A),~~~~~\{I_B\}=(1,2,3,...,B).
$$
Because it can be represented by a graph $G$ or by the corresponding
matrix $\underline{\underline{D}}$ we will consider the term as a graph
function $\mbox{\LARGE g}(\underline{\underline{D}})$. The scattering
amplitude $\Gamma_{AB}$ now can be re-written as
\begin{equation}                                 %**************eq6
\Gamma(\vec b)=\sum_\mu H_\mu \cdot \mbox{\LARGE g} \left(
\underline{\underline{D_\mu}}\right),
\label{eq6}
\end{equation}                                   %**************eq6
where summation runs on the set of all nonisomorphic graphs. From the
graph theory we have that the combinatorial coefficient at the function
of graph $G_\mu$ with $l-$components, $k_1$belonging to one class of
isomorphism, $k_2$ to another class, etc.,
($l=k_1+k_2+\cdot\cdot\cdot+k_j$) is equal to
\begin{eqnarray}                                 %**************eq7
H_\mu&=&\frac{A!}{(m_1!)^{k_1}(m_2!)^{k_2}\cdot\cdot\cdot(m_j!)^{k_j}
(A-\sum_{i=1}^jm_ik_i)! } \times
\nonumber\\                                      %**************eq7
&&\frac{B!}{(n_1!)^{k_1}(n_2!)^{k_2}\cdot\cdot\cdot(n_j!)^{k_j}
(B-\sum_{i=1}^j n_ik_i)! } \prod_{i=1}^j \left( \frac{m_i!n_i!}{
\mbox{\LARGE s}(G_i) } \right),
\label{eq7}
\end{eqnarray}                                   %**************eq7
where $m_i$ and $n_i$ are the numbers of the points of the sets $A$ and
$B$, respectively, in the component belonging to the i-th class of
isomorphism, and $\mbox{\LARGE s}(G_i)$ is the number of symmetries of
this component.

In the case of the elastic $d~{}^4He$ scattering the
$\underline{\underline{D}}$'s matrices will be
\[\begin{array}{ccccc}
\left(\begin{array}{cccc} 1&0&0&0\\0&0&0&0 \end{array}\right) &
\left(\begin{array}{cccc} 1&1&0&0\\0&0&0&0 \end{array}\right) &
\left(\begin{array}{cccc} 1&0&0&0\\1&0&0&0 \end{array}\right) &
\left(\begin{array}{cccc} 1&0&0&0\\0&1&0&0 \end{array}\right) &
\left(\begin{array}{cccc} 1&1&1&0\\0&0&0&0 \end{array}\right) \\
H_1=8      &  H_2=12    & H_3=4    & H_4=12      &  H_5=8     \\
\left(\begin{array}{cccc} 1&1&0&0\\1&0&0&0 \end{array}\right) &
\left(\begin{array}{cccc} 1&0&0&0\\1&0&1&0 \end{array}\right) &
\left(\begin{array}{cccc} 1&1&1&1\\0&0&0&0 \end{array}\right) &
\left(\begin{array}{cccc} 1&1&1&0\\1&0&0&0 \end{array}\right) &
\left(\begin{array}{cccc} 1&1&1&0\\0&0&0&1 \end{array}\right) \\
H_6=24     &  H_7=24    & H_8=2    & H_9=24      &  H_{10}=8  \\
\left(\begin{array}{cccc} 1&1&0&0\\1&1&0&0 \end{array}\right) &
\left(\begin{array}{cccc} 1&1&0&0\\1&0&1&0 \end{array}\right) &
\left(\begin{array}{cccc} 1&1&0&0\\0&0&1&1 \end{array}\right) &
\left(\begin{array}{cccc} 0&0&0&1\\1&1&1&1 \end{array}\right) &
\left(\begin{array}{cccc} 0&0&1&1\\0&1&1&1 \end{array}\right) \\
H_{11}=6   &  H_{12}=24 & H_{13}=6 & H_{14}=8    &  H_{15}=24 \\
\left(\begin{array}{cccc} 0&0&1&1\\1&1&0&1 \end{array}\right) &
\left(\begin{array}{cccc} 0&1&1&1\\1&0&1&1 \end{array}\right) &
\left(\begin{array}{cccc} 0&1&1&1\\0&1&1&1 \end{array}\right) &
\left(\begin{array}{cccc} 0&0&1&1\\1&1&1&1 \end{array}\right) &
\left(\begin{array}{cccc} 0&1&1&1\\1&1&1&1 \end{array}\right) \\
H_{16}=24  &  H_{17}=12 & H_{18}=4 & H_{19}=12    &  H_{20}=8 \\
\left(\begin{array}{cccc} 1&1&1&1\\1&1&1&1 \end{array}\right) &&&&\\
H_{21}=1   &            &        &           &
\end{array} \]

At $A=2$ and $B=4$ the amplitude $F_{AB}$ given by Eq.
(\ref{eq1}) is
\begin{equation}                                 %**************eq8
F_{24}(\vec q)=\frac{ip}{2\pi}\int d^2b~ e^{i\vec q \cdot \vec b}
\Gamma\left(\vec b \right),
\label{eq8}
\end{equation}                                   %**************eq8
\begin{equation}                                 %**************eq9
\Gamma\left(\vec b \right)= \langle\psi_d^i\psi_{He}^i |1-\prod_{j=1}^2
\prod_{k=1}^4 \left(1- \gamma(\vec b - \vec s_j+\vec
\tau_k)\right)|\psi_d^i\psi_{He}^i\rangle .
\label{eq9}
\end{equation}                                   %**************eq9
Introducing the distance between the nucleons in the deuteron, $\vec
r=(z,\vec s)$, and $\vec r_1=-\vec r_2=\vec r/2$ we have
\begin{equation}                                 %**************eq10
\Gamma\left(\vec b \right)= \langle\psi_d^i\psi_{He}^i |1-
\prod_{k=1}^4 \left(1- \gamma(\vec b - \frac{\vec s}{2}+\vec
\tau_k) \gamma(\vec b + \frac{\vec s}{2}+\vec \tau_k)\right)
|\psi_d^i\psi_{He}^i\rangle,
\label{eq10}
\end{equation}                                   %**************eq10
and
\begin{eqnarray}                                 %**************eq11
F_{24}(\vec q)&=&\frac{ip}{2\pi}\int d^2b~ e^{i\vec q \cdot \vec b}
\left[1-\prod_{k=1}^4 \left(1- \gamma(\vec b - \frac{\vec
s}{2}+\vec\tau_k)\right) \left(1- \gamma(\vec b + \frac{\vec
s}{2}+\vec\tau_k)\right)\right]\cdot
\nonumber\\                                      %**************eq11
&& ~~~~~~~~~~~~~~\left|\psi_d(\vec r) \right|^2 \left|\psi_{He}
\right|^2d^3r \prod_{k=1}^4 d^3\tau_k
\label{eq11}
\end{eqnarray}                                  %**************eq11
For the square module of $\psi_{d}$ we use the following
parametrization \cite{Azhgirei},
\begin{equation}                                %******************eq12
 {\left|{\psi}_d(\vec r)\right|}^{2}= \sum_{i=1}^{3} W_ie^{\frac{-\vec
r^2}{4\gamma_i}},
\label{eq12}
\end{equation}                                  %******************eq12
\begin{eqnarray}                                %******************eq
\gamma_{1}=225(GeV/c)^{2},&W_1=0.178/(4\pi\gamma_1)^{3/2},\nonumber\\
\gamma_{2}=45(GeV/c)^{2},&W_2=0.287/(4\pi \gamma_2)^{3/2},\nonumber\\
\gamma_{3}=25(GeV/c)^{2},&W_2=0.535/(4\pi \gamma_3)^{3/2}.\nonumber
%\label{eq}
\end{eqnarray}                                  %******************eq

The square module of $\psi_{He}$ was taken as \cite{Nik-Dakhno}.
\begin{equation}                                %******************eq13
{\left|{\psi}({\vec r}_1,\ldots,{\vec r}_4)\right|}^{2}= (2\pi)^3
\rho_c \delta\left(\sum_{i=1}^{4}{\vec r}_i\right)\prod_{i=1}^{4}
\varphi(\vec r_i).
\label{eq13}
\end{equation}                                  %******************eq13
We accept the two following parametrizations of $\varphi(\vec r)$
\begin{eqnarray}
(I)&& \varphi({\vec r})= exp[-{\vec r}^{2}/R_{1}^{2}],
%\label{parA}
\nonumber \\                                    %******************eq
%(B)&& \varphi({\vec r})= exp[-{\vec r}^{2}/R_{1}^{2}]-D_{1}exp[-{\vec
%r}^{2}/R_{2}^{2}],
%\label{parB}
%\nonumber \\                                    %******************eq
%(C)&& \varphi({\vec r})= (exp[-{\vec r}^{2}/2R_{1}^{2}]-D_{1}exp[-{\vec
%r}^{2}/2R_{2}^{2}])^{2},
%\label{parC}
%\nonumber \\                                    %******************eq
(II)&&\varphi({\vec r})= exp[-{\vec r}^{2}/R_{1}^{2}]+D_{1}exp[-{\vec
r}^{2}/R_{2}^{2}]-(1+D_{1}-D_{2}^{2})exp[-{\vec
r}^{2}/R_{3}^{2}]. \nonumber
\nonumber
\end{eqnarray}                                    %******************eq
The parameters are given in Table 1.
%------------------------------------------------------------------------
%---------------------------table1---------------------------------------
%------------------------------------------------------------------------
\begin{table} [h]                         %******************tabele1
\centering
\caption{Values of used parameters (from \protect{\cite{Nik-Dakhno}})}
\begin{tabular}{|c|c|c|c|c|c|} \hline
%    1   &    2      &     3     &     4     &   5   &   6   \\  \hline
         &  $R_1^2$  &  $R_2^2$  &  $R_3^2$  & $D_1$ & $D_2 $ \\ %\hline
         &$(GeV/c)^{-2}$&$(GeV/c)^{-2}$&$(GeV/c)^{-2}$& &  \\ \hline
    $I$  &  51.01   &          &          &       &         \\ \hline
    $II$ &  62.06   &   19.0   &   10.13  &  3.79 & 0.31     \\ \hline
\end{tabular}
\end{table}                                  %******************tabele1
%------------------------------------------------------------------------
%------------------------------------------------------------------------
%------------------------------------------------------------------------

\noindent We will use a general form for the function $\varphi$ as
\begin{equation}                                   %************eq14
\phi({\vec r})= \sum_{i=1}^{N}C_{i}e^{{\vec r}^2/R^{2}_{i}}.
\label{eq14}
\end{equation}                                     %************eq14

In the Eq. (\ref{eq13}) $\rho_c$ is the normalization constant given in
the Ref. \cite{Ahm-Uzh}. Substituting the Eqs. (\ref{eq12}),
(\ref{eq13}) and (\ref{eq14}) in the Eq. (\ref{eq11}) we have
\begin{eqnarray}                                 %**************eq15
F_{24}(\vec q)&=&\frac{ip}{2\pi}\rho_c\int d^2b~ e^{i\vec q \cdot \vec b}
\left[1-\prod_{j=1}^4 \left(1- \gamma(\vec b - \frac{\vec
s}{2}+\vec\tau_j)\right) \left(1- \gamma(\vec b + \frac{\vec
s}{2}+\vec\tau_j)\right)\right]
\nonumber\\                                      %**************eq15
&&\left(\sum_{j=1}^{3} W_je^{\frac{-\vec r^2}{4\gamma_j}}\right) d^3r
\prod_{j=1}^4 \left(\sum_{k=1}^{N}C_{k}e^{{\vec r}_j^2/R^{2}_{k}-
i\vec \alpha \vec r_j} \right) d^3r_j
\nonumber\\                                      %**************eq15
&=&\frac{ip\rho_c}{2\pi} \sum_{i_1=1}^{3}\sum_{i_j,j=2-5}^{N} W_{i_1}
C_{i_2} C_{i_3} C_{i_4} C_{i_5} \int d^2b~ e^{i\vec q \cdot \vec b}
\left[1-\prod_{k=1}^4 \left(1- \gamma(\vec b - \frac{\vec
s}{2}+\vec\tau_k)\right) \right.
\nonumber\\                                      %**************eq15
&& \left.\left(1- \gamma(\vec b + \frac{\vec s}{2}+\vec\tau_k)
\right)\right] e^{\frac{-\vec r^2}{4\gamma_{i_1}}}d^3r \prod_{k=2}^5
e^{{\vec r}_k^2/R^{2}_{i_j}-i \vec \alpha \vec r_k} d^3r_k
\label{eq15}
\end{eqnarray}                                  %**************eq15
where the integration on $\vec \alpha$ is performed in order to account
the $\delta$ function in the Eq. (\ref{eq13}) (see for details the Ref.
\cite{Ahm-Uzh}).

Using the graph functions, the Eq. (\ref{eq15}) can be written in
the form
\begin{eqnarray}                                 %**************eq16
F_{24}(\vec q)&=&\frac{ip\rho_c}{2\pi}
\sum_{i_1=1}^{3}\sum_{i_j,j=2-5}^{N} \sum_{\mu=1}^{21} H_\mu W_{i_1}
C_{i_2} C_{i_3} C_{i_4} C_{i_5} \int d^2b~ e^{i\vec q \cdot \vec b}
\cdot \mbox{\LARGE g} \left( \underline{\underline{D_\mu}}\right),
\nonumber\\                                      %**************eq16
&&~~~~~~~~~~~~~~~~~~~~~~~~~~~~ e^{\frac{-\vec r^2}{4\gamma_{i_1}}}d^3r
\prod_{k=1}^4 e^{{\vec r}_k^2/R^{2}_{i_j}} d^3\tau_k
\label{eq16}
\end{eqnarray}                                   %**************eq16

The result of the complete integration is \cite{Hansen}
\begin{equation}                                 %**************eq17
F_{24}(\vec q)=\frac{ip\rho_c}{2\pi} \sum_{n=1}^{21} \sum_{i_1=1}^{3}
\sum_{i_j,j=2-5}^{N} W_{i_1} C_{i_2} C_{i_3} C_{i_4} C_{i_5} (-1)^{n+1}
\beta^n \frac{\pi}{det|A|} exp\left[-\vec q^2
\frac{det|a|}{4det|A|}\right]
\label{eq17}
\end{equation}                                 %**************eq17
where
\begin{equation}                                 %**************eq18
det|\underline{\underline{A}}|=\left|\begin{array}{cc}
\underline{\underline{a}}&\underline{b^T} \\ \underline{b}&c
\end{array}\right|
\label{eq18}
\end{equation}                                 %**************eq18
$$
\underline{\underline{a}}=\left(\begin{array}{cccccc}
\frac{1}{R^2_{i_1}}+\frac{d_{11}+d_{21}}{2B_{NN}}&0&0&0&
\frac{-d_{11}+d_{21}}{2B_{NN}}&1\\
0&\frac{1}{R^2_{i_2}}+\frac{d_{12}+d_{22}}{2B_{NN}}&0&0&
\frac{-d_{12}+d_{22}}{2B_{NN}}&1\\
0&0&\frac{1}{R^2_{i_3}}+\frac{d_{13}+d_{23}}{2B_{NN}}&0&
\frac{-d_{13}+d_{23}}{2B_{NN}}&1\\
0&0&0&\frac{1}{R^2_{i_4}}+\frac{d_{14}+d_{24}}{2B_{NN}}&
\frac{-d_{14}+d_{24}}{2B_{NN}}&1\\
\frac{-d_{11}+d_{21}}{2B_{NN}}&\frac{-d_{12}+d_{22}}{2B_{NN}}&
\frac{-d_{13}+d_{23}}{2B_{NN}}&\frac{-d_{14}+d_{24}}{2B_{NN}}&
\frac{1}{4\gamma_i}+\frac{n}{8B_{NN}}&0\\
1&1&1&1&0&0
\end{array}\right)
$$
$$
\underline{b}=\left(\begin{array}{cccccc}
-\frac{d_{11}+d_{21}}{2B_{NN}}&-\frac{d_{12}+d_{22}}{2B_{NN}}&
-\frac{d_{13}+d_{23}}{2B_{NN}}&-\frac{d_{14}+d_{24}}{2B_{NN}}&
\frac{d_{11}+d_{12}+d_{13}+d_{14}-d_{21}-d_{22}-d_{23}-d_{24}}{2B_{NN}}&
\frac{n}{2B_{NN}}
\end{array}\right)
$$

To consider the elastic scattering at small momentum transfer we take
into account the Coulomb interaction (see the Ref. \cite{Avde}) writing
\begin{equation}                                 %**************eq19
\frac{d\sigma}{dt}=\frac{\pi}{p^2}\left|-n\frac{2p}{|\vec q|}
G(\vec q^2)e^{i\varphi}+F_N(\vec q)\right|^2,
\label{eq19}
\end{equation}                                 %**************eq19
where p is the laboratory momentum of the ${}^4He$, $n=Z_{He} \cdot
Z_{d}/137\beta$, $Z_{He}$, $Z_{d}$ are the charge of $He$ and $d$
respectively, $\beta=\frac{\nu}{c}$ is the ${}^4He$ velocity in the
laboratory system
\begin{equation}                                 %**************eq20
\phi=2n \ln(1.06/a|\vec q|) ; G(\vec q^2)=G_{He}(\vec q^2)
\cdot G_{d}(\vec \frac{q^2}{4}).
\label{eq20}
\end{equation}                                   %**************eq20
$G_{He}(t)$ is the ${}^4He$ form factor, and $G_{d}(t)$
is the deutron form factor,
$$G_{He}(t)=e^{11.9 \cdot t},$$
$$G_{d}(\frac{t}{4})=0.34e^{141.5\frac{t}{4}}+
0.58e^{26.1\frac{t}{4}}+0.08e^{15.5\frac{t}{4}}.$$

To perform the calculations one needs the $NN$ elastic
scattering amplitude parameters. A large number of publications both
theoretical and experimental have been devoted to processes the values
of the parameters. We take the values of the parameters --
$\sigma_{NN}^{tot}$, $\alpha_{NN}$ and $B_{NN}$ as follows:
$\sigma_{NN}^{tot}$ and $\alpha_{NN}$ were taken from the compilations
\cite{CERN} and \cite{UCRL}, respectively.  $B_{NN}$ was estimated
\cite{Ahm-Uzh} using $\sigma_{NN}^{tot}$ and $\sigma_{NN}^{el}$ taken
form the compilation \cite{CERN}. We call the set of the parameters as
set $A$.  Other set of the values (set $B$) was taken from the
paper by V.V.Avdeichikov \cite{Avde}, where the parameters were
obtained at fitting the experimental data on ${}^4He~d$ elastic
scattering.
%-----------------------------------------------------------------------
%-----------------------------Fig2--------------------------------------
%-----------------------------------------------------------------------
\begin{figure}[cbth]
\begin{center}
\psfig{file=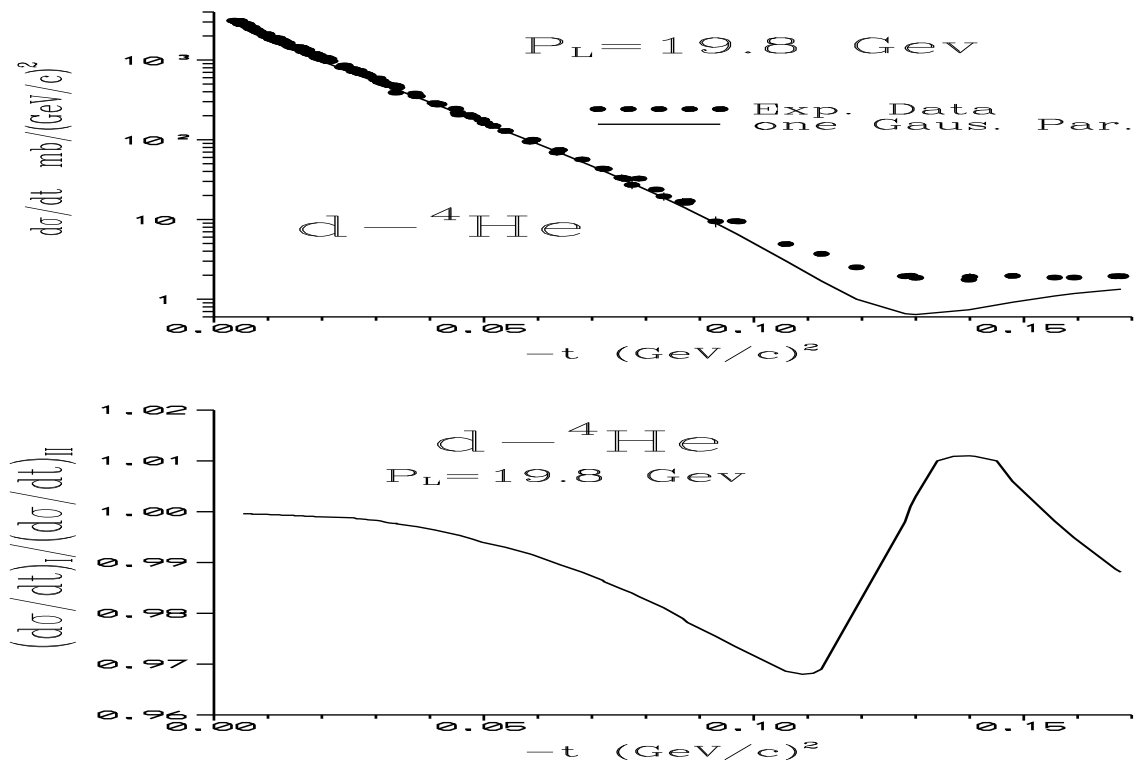,width=150mm,height=100mm,angle=0}
\caption{The ${}^4He~d$ differential cross sections at $P_l=19.8 GeV/c$.
Points are the experimental data \protect{\cite{Avde}}, lines
are our calculations using the set of values $A$.}
\end{center}
\end{figure}
%-----------------------------------------------------------------------
%-----------------------------------------------------------------------
%-----------------------------------------------------------------------

At the begining let us study the role of the different
parametrizations of the modules of the wave function of ${}^4He$.
The top figure 2 gives the ${}^4He~d$ elastic scattering differential
cross section calculated with the simplest parametrization I and set A.
The bottom figure presents the ratio of two calculations with
parametrization II and I. As seen, the difference between the
calculations is at the lavel of $3\%$. It is much smaller than the
difference between the experimental data and the calculations. Thus in
the following we will use only the parametrization I for simplicity.

The differential cross sections calculated with two sets of the $NN$
amplitude parameters $A$ and $B$ in a comparison with the
experimental data \cite{Avde} are shown in figure 3. As one can see,
the set $B$ reproduces the data quite well. At the same time it
is failed to describe the other reactions like $p~{}^4He$ scattering.
The set $A$ leads to a big disagreement with the experimental data
especially at large values of the momentum transfer $t$.  So, the main
problem of the analysis is a choice of the $NN$ parameters.
%-----------------------------------------------------------------------
%-----------------------------Fig3--------------------------------------
%-----------------------------------------------------------------------
\begin{figure}[cbth]
\begin{center}
\psfig{file=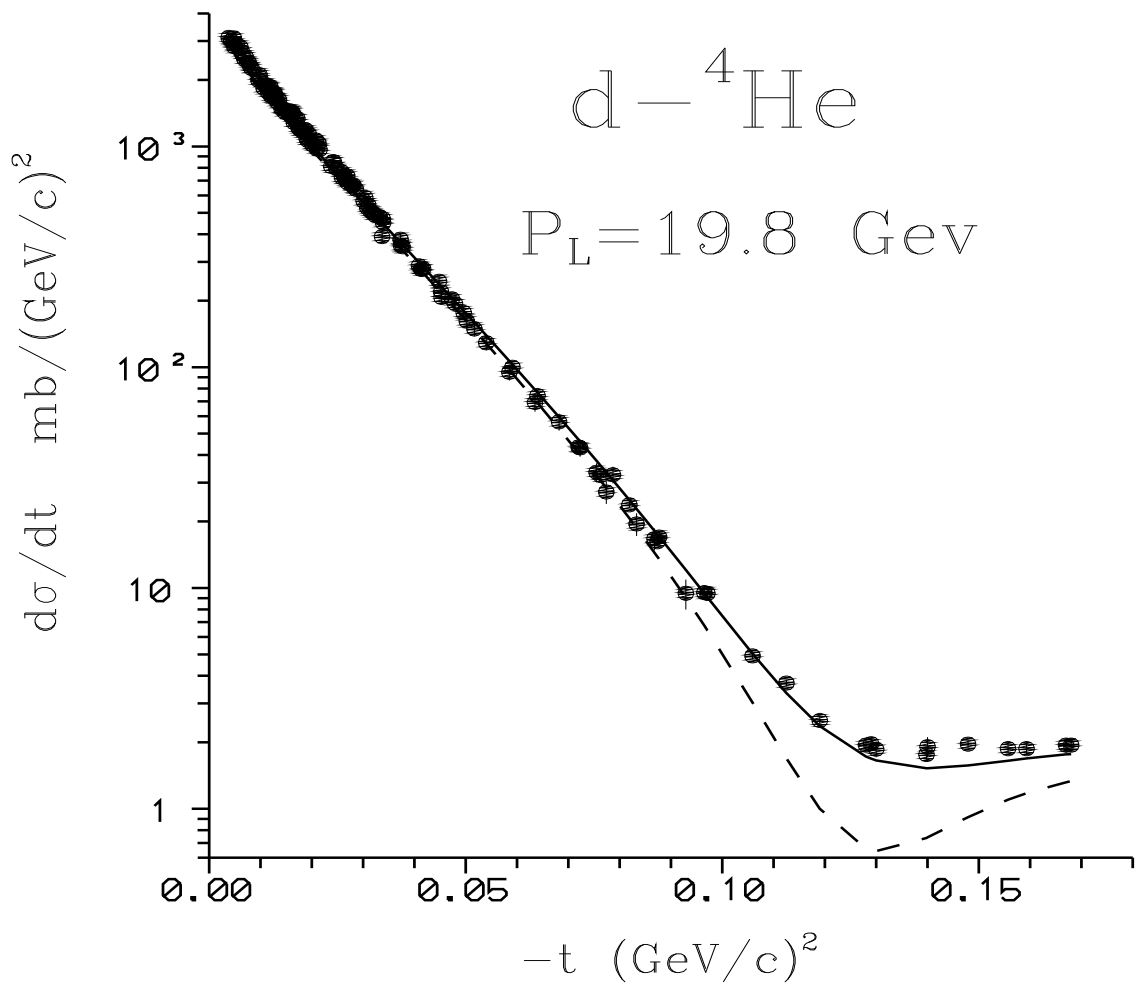,width=150mm,height=130mm,angle=0}
\caption{The ${}^4He~d$ differential cross sections at $P_l=19.8 GeV/c$.
Points are the experimental data \protect{\cite{Avde}}. Solid and
dashed lines are our calculations using the set of values $B$ and $A$,
respectively.}
\end{center}
\end{figure}
%-----------------------------------------------------------------------
%-----------------------------------------------------------------------
%-----------------------------------------------------------------------

In Ref. \cite{Azhgirei} the deutron - deutron scattering at momentum
8.9 GeV/$c$ was analyzed. The authors used two sets of the
NN-parameters presented in the table 2 as C and D sets. Both sets
allowed to describe the $dd$-scattering quite well (see Ref.
\cite{Azhgirei}).
%------------------------------------------------------------------------
%---------------------------table2---------------------------------------
%------------------------------------------------------------------------
\begin{table}[h]                        %******************tabele2
\centering
\caption{ The $NN-$amplitude parameter sets used in the calculations}
\vspace{5mm}
%\label{tabl1}
\begin{tabular}{|c|c|c|c|c|c|c|} \hline
%  1  &  2    &   3   &   4   &   5   &  6  & 7   \\
      &$\sigma_{NN}^{el}$& $\sigma_{PP}^t$ & $\sigma_{NP}^t$ &
      $\sigma_{NN}^t$ & $B_{NN}$&$\alpha_{NN}$    \\
      & $mb$  & $mb$  & $mb$    & $mb$  & $(GeV/c)^{-2}$& \\ \hline

  $A$ & 12& 41.67 & 42.04 & 41.86 & 7.39 & -0.33  \\ \hline
  $B$ &   &       &       & 41    & 7.9  & -0.55  \\ \hline
  $C$ &   &       &       & 42.4  & 6.3  & -0.43  \\ \hline
  $D$ &   &       &       & 42.4  & 7.3  & -0.43  \\ \hline
\end{tabular}
\end{table}                                  %******************tabele2
%-----------------------------------------------------------------------
%-----------------------------------------------------------------------
%-----------------------------------------------------------------------

As seen from the Fig. 4 both sets give a satisfactory description of
the ${}^4He~p$ elastic scattering except the region of large $t$.
\vspace{2mm}
%-----------------------------------------------------------------------
%-----------------------------Fig4--------------------------------------
%-----------------------------------------------------------------------
\begin{figure}[cbth]
\begin{center}
\psfig{file=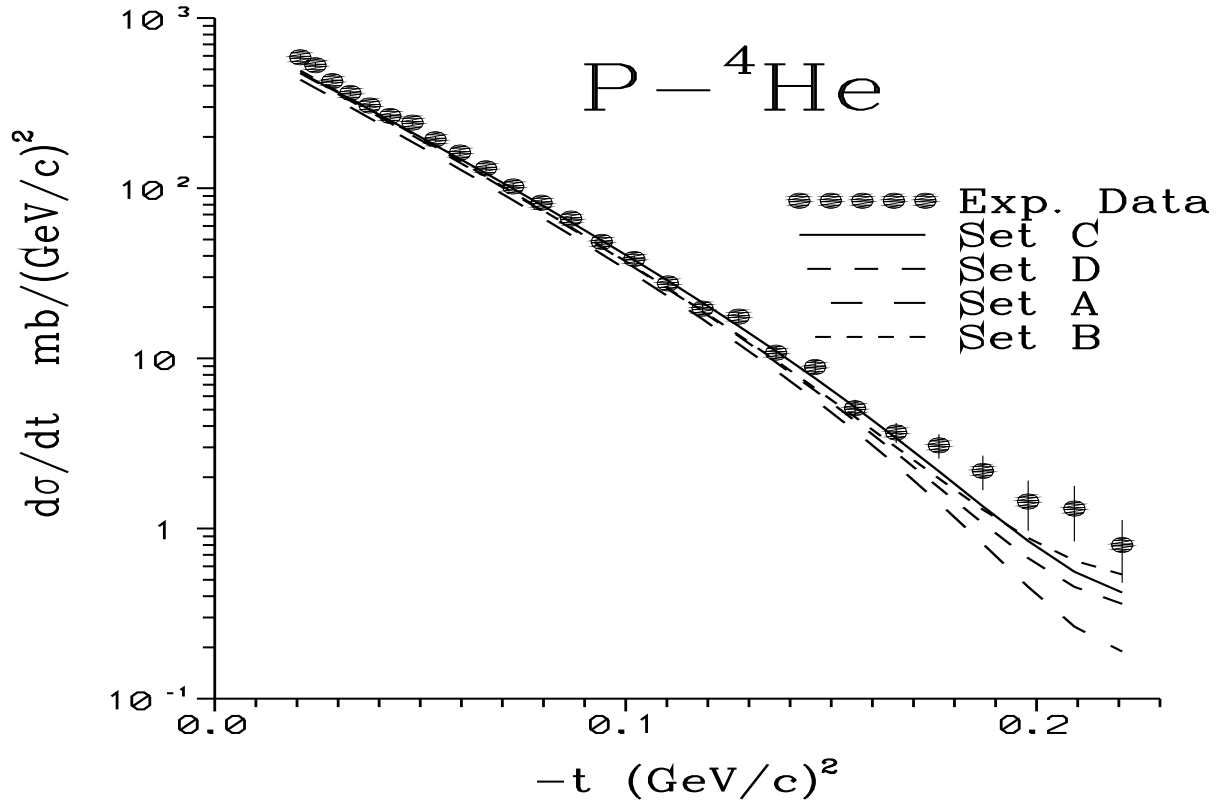,width=160mm,height=106mm,angle=0} % 107
\caption{The differential ${}^4He~p$ elastic scattering at $P_L=17.8$
GeV/$c$. Points are the experimental data \protect{\cite{Ableev}}.
Lines are our calculations with the different sets of $NN$-amplitude
parameters.}
\end{center}
\end{figure}
%-----------------------------------------------------------------------
%-----------------------------------------------------------------------
%-----------------------------------------------------------------------

The differential cross sections of the ${}^4He~d$ elastic scattering
calculated with the sets $C$ and $D$ are shown in Fig. 5 in a
comparison with the experimental data \cite{Avde}. As seen, we have a
big disagreement with the data in the region of the diffraction minimum
and also in the region of small momentum transfer $t$ with the set $D$.
Inverse situation takes place with the set $C$, the calculation are
above the experimental data at small $t$ and above the data at large
$t$.
%-----------------------------------------------------------------------
%-----------------------------Fig5--------------------------------------
%-----------------------------------------------------------------------
\begin{figure}[cbth]
\begin{center}
\psfig{file=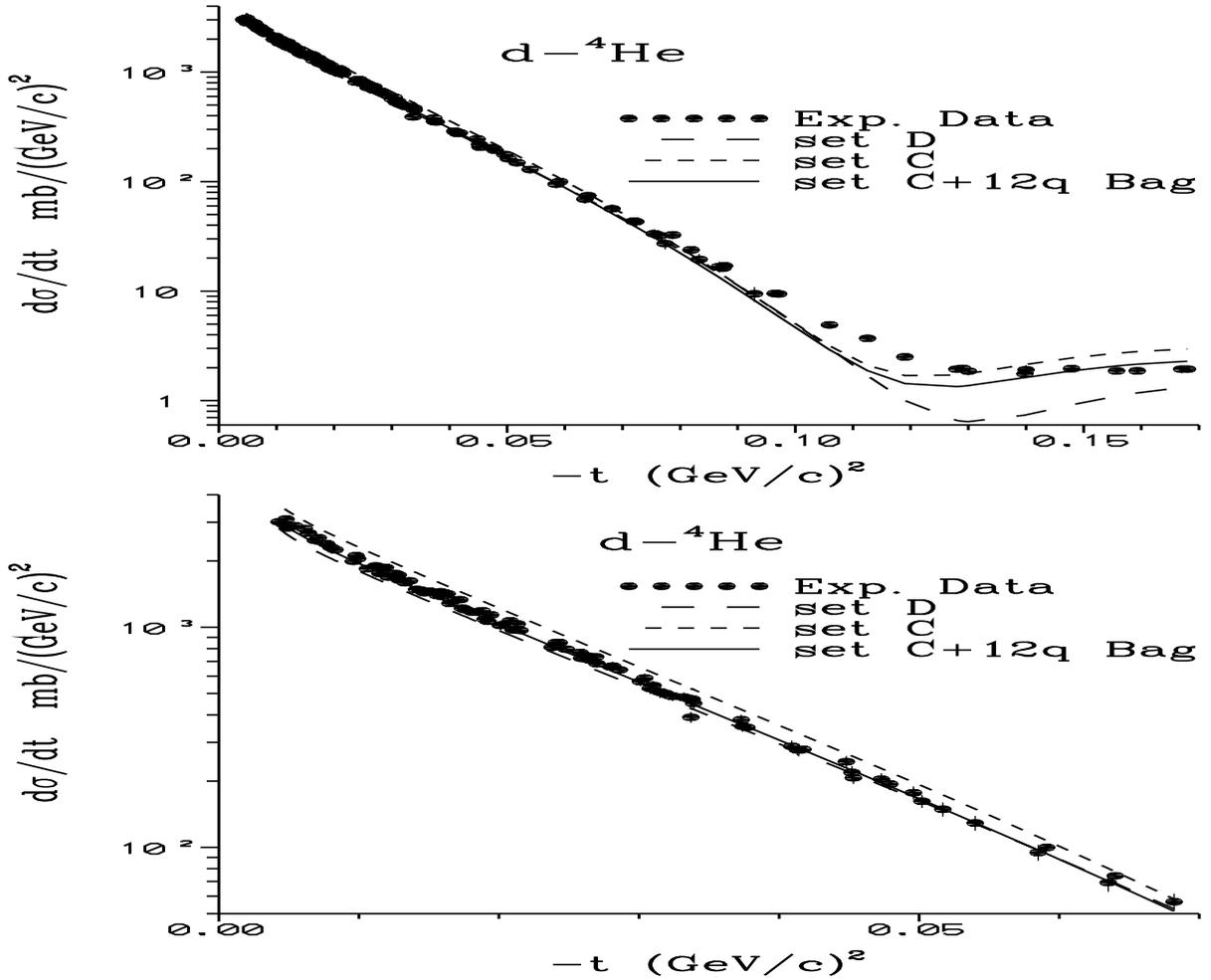,width=160mm,height=130mm,angle=0}
\caption{The differential ${}^4He~d$ elastic scattering at $P_L=19.8$
GeV/$c$. Points are the experimental data \protect{\cite{Avde}}.
Short dasahed and long dashed lines are our calculations using the
sets $C$ and $D$, respectively. Solid lines -- calculations
performed with inclusion of the 12-quark bag state of ${}^4He$ and with
the parameter set C.}
\end{center}
\end{figure}
%-----------------------------------------------------------------------
%-----------------------------------------------------------------------
%-----------------------------------------------------------------------

Summing up we conclude that it is impossible to describe simultaneously
the $p~{}^4He$ and $d~{}^4He$ elastic scattering cross sections using
the same set of the $NN$-amplitude parameters. The parameters what are
quite well for understanding $dd$-scattering \cite{Azhgirei} can not be
applied in the case of $d~{}^4He$ scattering.

%**********************************************************************
\section{Manifestation of the twelve quark bag admixture}
%*********************************************************************
Let us neglect all transition amplitudes like that $|4N>~ \rightarrow
|12q>$, $|12q>~ \rightarrow |4N>$ following papers \cite{Nik-Dakhno,
Ahm-Uzh}. In this case the $d~{}^4He$ scattering amplitude will be
\begin{equation}                                 %******************21
F_{d~{}^4He}=\left(1-w_{12q}\right)F_{d,4N} + w_{12q}F_{d,12q}
\label{eq21}
\end{equation}                                   %******************21
where $F_{d,4N}$ is the Glauber amplitude of the deutron - four nucleon
scattering given by the Eq. (\ref{eq11}), $F_{d,12q}$ is the deutron -
$12q$ bag scattering amplitude, and $w_{12q}$ is the weight
of the $12q$ bag quark state in the ground state of the ${}^4He$
nucleus. The values of the $12q$ bag weight was estimated in our previous
paper \cite{Ahm-Uzh} as $w_{12q}=10.5\%$.

$F_{d,12q}$ amplitude can be estimated as $d~p$ one.
$$                                %******************22
F_{d,12q}=\frac{ip}{2\pi}\sum_{j=1}^3 W_j \int d^2b e^{i\vec q \cdot
\vec b} \left[ \gamma_{N,12q}(\vec b-\frac{\vec
s}{2})+\gamma_{N,12q}(\vec b+\frac{\vec s}{2}) -\right.
$$
\begin{equation}
\left. \gamma_{N,12q}(\vec
b-\frac{\vec s}{2}) \gamma_{N,12q}(\vec b+\frac{\vec s}{2})\right]
e^{\frac{-\vec r^2}{4\gamma_j}} d^3r
\label{eq22}
\end{equation}                                   %******************22
Using the gaussian parametrization for the nucleon - 12q bag scattering
amplitude
\begin{equation}                                 %******************23
\gamma_{N,12q}(\vec b)=\frac{\sigma_{12q}^t}{4\pi B_{12q} } \cdot
e^{\frac{-\vec b^2}{2 B_{12q} } },
\label{eq23}
\end{equation}                                   %******************23
all intergations in the Eq. (\ref{eq22}) can be done exactly. In
particular,
\begin{eqnarray}                                 %******************24
F_{d,12q}^1&=&\int d^2b e^{i\vec q \cdot \vec b}
\gamma_{N,12q}(\vec b-\frac{\vec s}{2}) e^{\frac{-\vec r^2}{4\gamma_j}}
d^3r
\nonumber\\                                      %**************eq24
&=& \frac{\sigma_{12q}^t}{4\pi B_{12q}} \int d^2b e^{i\vec q \cdot \vec
b} e^{\frac{-\left(\vec b-\frac{\vec s}{2}\right)^2}{2B_{12q}}}
e^{\frac{-\vec r^2}{4\gamma_j}}d^2sdz
\nonumber\\                                      %**************eq24
&=& \frac{\sigma_{12q}^t}{4\pi B_{12q}} \int d^2b e^{i\vec q \cdot \vec
b} exp\left[-\left(\frac{2B_{12q}+\gamma_j}{8B_{12q}\gamma_j} \right)
\left(s^2- \frac{2\gamma_j}{2B_{12q}+\gamma_j} \vec b \right)^2\right]
\nonumber\\                                      %**************eq24
&&exp\left[-\frac{\vec b^2}{2 B_{12q}+\gamma_j}\right]
e^{\frac{-\vec z^2}{4\gamma_j}} d^2s dz
\nonumber\\                                      %**************eq24
&=&\frac{\sigma_{12q}^t}{4\pi B_{12q}}
\left(\frac{8B_{12q}\gamma_j\pi}{2B_{12q}+\gamma_j}\right)
\left(4\gamma_j\pi\right)^{\frac{1}{2}}
\int d^2b e^{i\vec q \cdot \vec b}
exp\left[-\frac{\vec b^2}{2B_{12q}+\gamma_j}\right]
\nonumber\\                                      %**************eq24
&=&\frac{\sigma_{12q}^t}{4\pi B_{12q}}
\left(\frac{8B_{12q}\gamma_j\pi}{2B_{12q}+\gamma_j}\right)
\left(4\gamma_j\pi\right)^{\frac{1}{2}}
\left(2B_{12q}+\gamma_j\right)\pi
exp\left[-\left(\frac{2B_{12q}+\gamma_j}{4}\right)\vec q^2\right]
\nonumber\\                                      %**************eq24
&=&2 \pi \gamma_j \sigma_{12q}^t \left(4\gamma_j\pi\right)^{\frac{1}{2}}
exp\left[-\left(\frac{2B_{12q}+\gamma_j}{4}\right)\vec q^2\right].
\label{eq24}
\end{eqnarray}                                   %******************24

The second is given as
\begin{eqnarray}                                 %******************25
F_{d,12q}^2&=&\int d^2b e^{i\vec q \cdot \vec b}
\gamma_{N,12q}(\vec b-\frac{\vec s}{2}) \gamma_{N,12q}(\vec
b+\frac{\vec s}{2}) e^{\frac{-\vec r^2}{4\gamma_j}} d^3r
\nonumber\\                                      %**************eq25
&=& \left(\frac{\sigma_{12q}^t}{4\pi B_{12q}}\right)^2 \int d^2b
e^{i\vec q \cdot \vec b}
e^{\frac{-\left(\vec b- \frac{\vec s}{2} \right)^2}{2B_{12q}}}
e^{\frac{-\left(\vec b+ \frac{\vec s}{2} \right)^2}{2B_{12q}}}
e^{\frac{-\vec r^2}{4\gamma_j}}d^2sdz
\nonumber\\                                      %**************eq25
&=& \left(\frac{\sigma_{12q}^t}{4\pi B_{12q}}\right)^2 \int d^2b
e^{i\vec q \cdot \vec b} e^{-\frac{\vec b^2}{B_{12q}}}
exp\left[-\left(\frac{B_{12q}+\gamma_j}{ 4B_{12q} \gamma_j}\right)
\vec s^2\right] e^{\frac{-\vec z^2}{4\gamma_j}}d^2sdz
\nonumber\\                                      %**************eq25
&=&\left(\frac{\sigma_{12q}^t}{4\pi B_{12q}}\right)^2
\left(4\gamma_j\pi\right)^{\frac{1}{2}}
\left(\frac{4B_{12q} \gamma_j \pi}{B_{12q}+\gamma_j}\right)
( B_{12q} \pi) e^{-\frac{B_{12q}}{4}\vec q^2}
\nonumber\\                                      %**************eq25
&=&\frac{\left(\sigma_{12q}^t\right)^2 \gamma_j}{4\pi
\left(B_{12q}+\gamma_j\right)} e^{-\frac{B_{12q}}{4}\vec q^2}
\label{eq25}
\end{eqnarray}                                   %******************25
The $d~12q$ bag amplitude will be
\begin{equation}                                 %******************30
F_{d,12Q}=\frac{ip}{2\pi}\sum_{i=1}^3W_i \left(4\pi \gamma_i
\right)^{\frac{1}{2}} \left[ 2\pi \gamma_i \sigma_{12q}^t
e^{-\left(\frac{2B_{12q}+\gamma_i}{4} \right)q^2} -\frac{
(\sigma_{12q}^t)^2 \gamma_i}{ 4(B_{12q}+\gamma_i)}
e^{-\frac{B_{12q}}{4}q^2}\right].
\label{eq30}
\end{equation}                                   %******************30
The values of the nucleon-$12q$ bag amplitude parameters were
estimated in the Ref. \cite{Ahm-Uzh}:  $w_{12q}=10.5\%$,
$\sigma_{12q}^t=34 mb$ and $B_{12q}= 23 \left( GeV \right)^{-2}$. These
values have been used with the parameter set $C$ for the calculations
of the differential cross section with and without the $12q$ bag
admixture shown in the Fig. 5. One can see that when the $12q$ bag
admixture is included in the calculations it gives a good description
of the data at small  and large values of $t$. The discrepancy between
the calculations and the experimental data in the region 0.09 $\leq |t|
\leq$ 0.13 $(GeV/c)^{-2}$ can be erased at taking into account the
D-wave of the deutron \cite{Inoz}.

\section*{Conclusion}
\begin{enumerate}
\item
Within the framework of the Glauber theory it is impossible to describe
simultaneously the $p~{}^4He$ and $d~{}^4He$ elastic scattering cross
sections using the same set of the $NN$-amplitude parameters.

\item The 12q bag admixture to the ground state of the ${}^4He$ nucleus
manifests itself in the $d~{}^4He$ elastic scattering in all region of
the momentum transfer. At small $t$ the effect can be at the level of
$\sim$ 10 \%. At large $t$ it can be $\sim$ 30 \%.

\item
To study the effect at an experiment it is needed to measure the
$d~{}^4He$ elastic cross section with absolute normalization accuracy
better than 10 \%.

\end{enumerate}

V.V.Uzhinskii thanks RFBR (grand No ~00-01-00307,  01-02-16431)
and INTAS (grand No ~00-00366) for their financial support.
A.M. Mosallem is thankful to Profs. A.B.B. Kalil and K.M. Hanna
for support, and  JINR officials for hospitality.

\end{document}